\documentclass[conference]{IEEEtran}
\IEEEoverridecommandlockouts
\usepackage{listings}
\usepackage{multicol}
\usepackage{lipsum}
\usepackage{mathtools}
\usepackage{amsthm}
\usepackage{acronym}
\usepackage{url}
\usepackage{stfloats}
\usepackage{amsfonts}
\usepackage{acronym}
\usepackage{cite}
\usepackage{multirow}
\usepackage{bm}
\usepackage{amsmath,amssymb}
\usepackage{graphicx}
\usepackage[table,xcdraw]{xcolor}
\usepackage[english]{babel}
\usepackage[caption=false,font=footnotesize]{subfig}
\usepackage[geometry]{ifsym}
\usepackage{array}
\usepackage{flushend}
\usepackage[utf8]{inputenc}
\usepackage[T1]{fontenc}
\usepackage{fancyhdr}

\usepackage{tikz}
\usepackage{lipsum}
\usetikzlibrary{calc,shadings,patterns}
\makeatletter
\tikzset{%
  remember picture with id/.style={%
    remember picture,
    overlay,
    save picture id=#1,
  },
  save picture id/.code={%
    \edef\pgf@temp{#1}%
    \immediate\write\pgfutil@auxout{%
      \noexpand\savepointas{\pgf@temp}{\pgfpictureid}}%
  },
  if picture id/.code args={#1#2#3}{%
    \@ifundefined{save@pt@#1}{%
      \pgfkeysalso{#3}%
    }{
      \pgfkeysalso{#2}%
    }
  }
}

\def\savepointas#1#2{%
  \expandafter\gdef\csname save@pt@#1\endcsname{#2}%
}

\def\tmk@labeldef#1,#2\@nil{%
  \def\tmk@label{#1}%
  \def\tmk@def{#2}%
}

\tikzdeclarecoordinatesystem{pic}{%
  \pgfutil@in@,{#1}%
  \ifpgfutil@in@%
    \tmk@labeldef#1\@nil
  \else
    \tmk@labeldef#1,(0pt,0pt)\@nil
  \fi
  \@ifundefined{save@pt@\tmk@label}{%
    \tikz@scan@one@point\pgfutil@firstofone\tmk@def
  }{%
  \pgfsys@getposition{\csname save@pt@\tmk@label\endcsname}\save@orig@pic%
  \pgfsys@getposition{\pgfpictureid}\save@this@pic%
  \pgf@process{\pgfpointorigin\save@this@pic}%
  \pgf@xa=\pgf@x
  \pgf@ya=\pgf@y
  \pgf@process{\pgfpointorigin\save@orig@pic}%
  \advance\pgf@x by -\pgf@xa
  \advance\pgf@y by -\pgf@ya
  }%
}

\makeatother

\newcounter{hatchNumber}
\usepackage{cuted}
\makeatletter
\newif\ifAC@uppercase@first%
\def\Aclp#1{\AC@uppercase@firsttrue\aclp{#1}\AC@uppercase@firstfalse}%
\def\AC@aclp#1{%
	\ifcsname fn@#1@PL\endcsname%
	\ifAC@uppercase@first%
	\expandafter\expandafter\expandafter\MakeUppercase\csname fn@#1@PL\endcsname%
	\else%
	\csname fn@#1@PL\endcsname%
	\fi%
	\else%
	\AC@acl{#1}s%
	\fi%
}%
\def\Acp#1{\AC@uppercase@firsttrue\acp{#1}\AC@uppercase@firstfalse}%
\def\AC@acp#1{%
	\ifcsname fn@#1@PL\endcsname%
	\ifAC@uppercase@first%
	\expandafter\expandafter\expandafter\MakeUppercase\csname fn@#1@PL\endcsname%
	\else%
	\csname fn@#1@PL\endcsname%
	\fi%
	\else%
	\AC@ac{#1}s%
	\fi%
}%
\def\Acfp#1{\AC@uppercase@firsttrue\acfp{#1}\AC@uppercase@firstfalse}%
\def\AC@acfp#1{%
	\ifcsname fn@#1@PL\endcsname%
	\ifAC@uppercase@first%
	\expandafter\expandafter\expandafter\MakeUppercase\csname fn@#1@PL\endcsname%
	\else%
	\csname fn@#1@PL\endcsname%
	\fi%
	\else%
	\AC@acf{#1}s%
	\fi%
}%
\def\Acsp#1{\AC@uppercase@firsttrue\acsp{#1}\AC@uppercase@firstfalse}%
\def\AC@acsp#1{%
	\ifcsname fn@#1@PL\endcsname%
	\ifAC@uppercase@first%
	\expandafter\expandafter\expandafter\MakeUppercase\csname fn@#1@PL\endcsname%
	\else%
	\csname fn@#1@PL\endcsname%
	\fi%
	\else%
	\AC@acs{#1}s%
	\fi%
}%
\edef\AC@uppercase@write{\string\ifAC@uppercase@first\string\expandafter\string\MakeUppercase\string\fi\space}%
\def\AC@acrodef#1[#2]#3{%
	\@bsphack%
	\protected@write\@auxout{}{%
		\string\newacro{#1}[#2]{\AC@uppercase@write #3}%
	}\@esphack%
}%
\def\Acl#1{\AC@uppercase@firsttrue\acl{#1}\AC@uppercase@firstfalse}
\def\Acf#1{\AC@uppercase@firsttrue\acf{#1}\AC@uppercase@firstfalse}
\def\Ac#1{\AC@uppercase@firsttrue\ac{#1}\AC@uppercase@firstfalse}
\def\Acs#1{\AC@uppercase@firsttrue\acs{#1}\AC@uppercase@firstfalse}

\acrodef{SIC}{successive interference cancellation}
\acrodef{PAPR}{peak-to-average-power ratio}
\acrodef{APAC}{aperiodic autocorrelation}
\acrodef{OFDM}{orthogonal frequency division multiplexing}
\acrodef{DFT}{discrete Fourier transform}
\acrodef{DC}{direct current}
\acrodef{CS}{complementary sequence}
\acrodef{GCP}{Golay complementary pair}
\acrodef{ANF}{algebraic normal form}
\acrodef{PSK}{phase shift keying}
\acrodef{QAM}{quadrature amplitude modulation}
\acrodef{QPSK}{quadrature phase shift keying}
\acrodef{GDJ}{Golay-Davis-Jedwab}
\acrodef{PMEPR}{peak-to-mean envelope power ratios}
\acrodef{FFT}{fast Fourier transform}
\acrodef{BER}{bit-error ratio}
\acrodef{SNR}{signal-to-noise ratio}
\acrodef{4G}{fourth generation}
\acrodef{5G}{fifth generation}
\acrodef{ADAS}{advanced driver assistance system }

\acrodef{NR}{new radio}
\acrodef{LTE}{long-term evolution}
\acrodef{PTS}{partial transmit sequences}
\acrodef{PSD}{power spectral density}
\acrodef{LDPC}{low-density parity check}
\acrodef{SE}{spectral efficiency}
\acrodef{eLAA}{enhanced licensed-assisted access}
\acrodef{NR-U}{NR-unlicensed}
\acrodef{RM}{Reed-Muller}
\acrodef{NP-hard}{Non-deterministic Polynomial-time hardness}
\acrodef{AE}{autoencoder}
\acrodef{DNN}{deep neural network}
\acrodef{OFDM-AE}{OFDM-based autoencoder}
\acrodef{CP}{cyclic prefix}
\acrodef{AWGN}{additive white Gaussian noise}
\acrodef{P2C}{polar-to-Cartesian}
\acrodef{CFR}{channel frequency response}
\acrodef{ReLU}{rectified linear unit}
\acrodef{MMSE}{minimum mean square error}
\acrodef{BPSK}{binary phase shift keying}
\acrodef{BLER}{block error rate}
\acrodef{ML}{machine learning}
\acrodef{PHY}{physical layer}
\acrodef{PA}{power amplifier}
\acrodef{IDFT}{inverse DFT}
\acrodef{DoF}{degrees-of-freedom}
\acrodef{IoT}{Internet-of-Things}
\acrodef{M2M}{machine-to-machine}

\acrodef{QZSS}{Quasi-Zenith Satellite System}
\acrodef{GALILEO}{Galileo Satellite Navigation}
\acrodef{GPS}{Global Positioning System}
\acrodef{GLONASS}{Globalnaya Navigatsionnaya Sputnikovaya Sistema}

\acrodef{DFT-s-OFDM}{discrete Fourier transform-spread orthogonal frequency division multiplexing}
\acrodef{MMSE}{minimum mean square error}
\acrodef{FDE}{frequency-domain equalization}
\acrodef{FrFT}{fractional Fourier transform}
\acrodef{TF}{time-frequency}
\acrodef{BFSK}{binary frequency-shift keying}
\acrodef{CSS}{chirp spread spectrum}
\acrodef{BCSS}{binary chirp spread spectrum}
\acrodef{EVA}{extended vehicular A}
\acrodef{MIMO}{multi-input multi-output}
\acrodef{LoRa}{long range}
\acrodef{HF}{high-frequency}
\acrodef{FDSS}{frequency-domain spectral shaping}
\acrodef{UAM}{urban air mobility}
\acrodef{CNS}{communication, navigation and surveillance}
\acrodef{ATM}{air traffic management}
\acrodef{ATC}{air traffic control}
\acrodef{C2}{command and control}
\acrodef{nC2}{non-command and control}
\acrodef{AI}{artificial intelligence}
\acrodef{NTN}{non-terrestrial networks}
\acrodef{UAS}{unmanned aircraft systems}
\acrodef{UTM}{UAS traffic management}
\acrodef{VFR}{visual flight rules} 
\acrodef{IFR}{instrument flight rules} 
\acrodef{RPIC}{remote pilot in command}
\acrodef{DAA}{detect \& avoid}
\acrodef{LEO}{low-earth orbit}
\acrodef{MEO}{medium-earth orbit}
\acrodef{GEO}{geosynchronous-earth orbit}
\acrodef{SWAP}{size, weight and power}
\acrodef{CNPC}{control and non-payload communication} 
\acrodef{Con-Ops}{concept of operations}
\acrodef{RN}{relay node}
\acrodef{LOS}{line-of-sight}
\acrodef{NLOS}{non-line-of-sight}
\acrodef{BLOS}{beyond line-of-sight}
\acrodef{HAPS}{high-altitude platforms}
\acrodef{PTRS}{phase-tracking reference symbols}
\acrodef{UE}{user equipment}
\acrodef{GSO}{geostationary synchronous orbit}
\acrodef{GNSS}{global navigation satellite system}
\acrodef{NGSO}{non-geostationary synchronous orbit}
\acrodef{SC}{single carrier}
\acrodef{SC-FDMA}{single carrier frequency division multiple access}
\acrodef{OBO}{out-of-band emission}
\acrodef{UL}{uplink}
\acrodef{DL}{downlink}
\acrodef{VDLM2}{VHF data link mode 2}
\acrodef{PA}{power amplifier}
\acrodef{PAPR}{peak-to-average ratio}
\acrodef{RTCA}{radio technical commission for aeronautics}
\acrodef{mmWave}{millimeter wave}
\acrodef{BLOS}{beyond line-of-sight}
\acrodef{ATSP}{asymmetric traveling salesman problem}
\acrodef{RX}{receiver}
\acrodef{TX}{transmitter}
\acrodef{UAV}{unmanned air vehicle}
\acrodef{RTK}{real-time kinematic}
\acrodef{VLL}{very low-level}
\acrodef{GS}{ground station}
\acrodef{TDD}{time-domain duplexing}
\acrodef{LIDAR}{light, imaging detection, and ranging}
\acrodef{FMCW}{frequency modulated continuous wave}
\acrodef{ADS-B}{automatic dependent surveillance-broadcast}
\acrodef{FAA}{federal aviation administration}
\acrodef{EW}{electronic warfare}
\acrodef{VDL}{VHF digital link}
\acrodef{3GPP}{3rd generation partnership project}
\acrodef{LDACS}{L-band digital aeronautical communication system}
\acrodef{D2D}{device-to-device}
\acrodef{AeroMACS}{aeronautical mobile airport communication system}
\acrodef{FEC}{forward error correction}
\acrodef{RCP}{raised cosine pulse}
\acrodef{IR}{impulse radio}
\acrodef{FBMC}{filter-bank multicarrier}
\acrodef{UFMC}{universal filtered multicarrier}
\acrodef{IMU}{Inertial Measurement Unit}
\acrodef{TSP}{travelling salesman problem}
\acrodef{VTOL}{vertical takeoff and landing}
\acrodef{IFFT}{inverse fast Fourier transform}
\acrodef{ACAS}{airborne collision avoidance systems}
\acrodef{ARC}{aviation rulemaking committee}
\acrodef{SSR}{secondary surveillance radar}
\acrodef{HARQ}{hybrid automatic repeat request}
\acrodef{AG}{air-to-ground}

\def\BibTeX{{\rm B\kern-.05em{\sc i\kern-.025em b}\kern-.08em
    T\kern-.1667em\lower.7ex\hbox{E}\kern-.125emX}}
\begin{document}

\title{{\fontsize{24}{26}\selectfont{ Requirements and Technologies Towards UAM: Communication, Navigation, and Surveillance }}\fontsize{16}{18}\selectfont
}


\author{\IEEEauthorblockN{\textit{M. Cenk~Ert\"urk, Honeywell Aerospace, Redmond, WA.} \\ \textit{Nozhan Hosseini, Hosseinali Jamal, Alphan \c{S}ahin, David Matolak, University of South Carolina, Columbia, SC.} \\ \textit{Jamal Haque, University of South Florida, Tampa, FL.}}
}

\maketitle
\LARGE\textrm{Abstract}\\ \normalsize 

\Ac{UAM} is a concept for creating an airborne transportation system that operates in urban settings with an on-board pilot and/or \ac{RPIC}, or with a fully autonomous architecture. Although the passenger traffic will be mostly in and near urban environments, \ac{UAM} is also being considered for air cargo, perhaps between cities. Such capability is pushing the current \ac{CNS} / \ac{ATM} systems that were not designed to support these types of aviation scenarios. The \ac{UAM} aircraft will be operating in a congested environment, where \ac{CNS} and \ac{ATM} systems need to provide integrity, robustness, security, and very high availability  for safety of \ac{UAM} operations while evolving. As \ac{UAM} is under research by academia and government agencies, the industry is driving technology towards aircraft prototypes. Critical \ac{UAM} requirements are derived from \ac{C2} (particularly for \ac{RPIC} scenario), data connectivity for passengers and flight systems, \ac{UAS} to \ac{UAS} communication to avoid collision, and data exchange for positioning and surveillance. In this paper, we study connectivity challenges and present requirements towards a robust \ac{UAM} architecture through its concept of operations. In addition, we review the existing/potential \ac{CNS} technologies towards \ac{UAM}, i.e., \ac{3GPP} \ac{5G} \ac{NR}, navigation \ac{DAA}, and satellite systems and present conclusions on a future road-map for \ac{UAM} \ac{CNS} architecture.

\textit{Keywords:} \ac{5G}, \ac{CNS}, \ac{DAA}, NASA, requirement analysis, \ac{RTCA}, satellite systems, \ac{UAM}.

\vspace{\baselineskip}

\Large\textrm{1- Introduction}\\ \normalsize

\pagestyle{fancy}
\fancyhead[CO,C]{\fontfamily{phv}\fontsize{7}{11}\selectfont \textcolor{blue}{ Citation: M. C. Erturk, N. Hosseini, H. Jamal, A. Sahin, D. Matolak and J. Haque, "Requirements and Technologies Towards UAM: Communication, Navigation, and Surveillance," 2020 Integrated Communications, Navigation, Surveillance Conference (ICNS), Herndon, VA, 2020.}}
\fancyfoot[LO,L]{\fontfamily{phv}\fontsize{7}{11}\selectfont  This paper is accepted to be published in IEEE ICNS USA 2020} 

NASA’s \ac{UAM} represents one of the most ambitious challenges for the aviation industry. With the goal of ultra-safe and efficient aircraft movement of passengers and freight near and within urban environments, this operation will require advances in a range of engineering domains, including propulsion, airframe design, \ac{ATM} and \ac{CNS}.

A NASA \ac{UAM} market study \cite{NASA_2019} defined three most challenging \ac{UAM} use cases as, last-mile delivery, air metro, and air taxi, for possible implementation by 2030. NASA’s \ac{UTM} program has been operating for a few years, progressively advancing use of \ac{UAS} operating at very low altitudes (e.g. below 400 feet, sometimes termed {\em \ac{VLL}} air-space/operations). \ac{UTM} is planned to provide a set of traffic management services via a federated group of \ac{UTM} Service Suppliers, comparable to traditional \ac{ATC} services provided to \ac{VFR} and \ac{IFR} aircraft.

Rapid delivery of packages ($<5$~lb) from a local depot hub to a customer or next depot represents so-called last mile delivery. At predetermined schedules, routes and stops throughout a city are envisioned, similar to bus or metro routes on the ground. An air taxi scheme (door to door) ride-sharing operation with customer defined drop-off and destination depots is another \ac{UAM} application. All three of these applications require sophisticated route planning and accurate navigation. An efficient \ac{DAA}  system is also required for \ac{UAM} operations and planning.

The focus of this paper is to understand the connectivity challenges and determine the requirements toward robust \ac{UAM} through its concept of operations. Once the baseline scenario and its requirements are established, we investigate the existing/potential CNS technologies towards \ac{UAM}, i.e., \ac{3GPP} \ac{5G} \ac{NR}, navigation \ac{DAA}, and satellite systems, and provide a potential road-map for \ac{UAM} CNS architecture.

The rest of this paper is organized as follows. In Section 2, we define requirements for \ac{UAM} scenarios that addresses the potential connectivity challenges. Section 3 presents the NASA and \ac{RTCA} activities in terms of programs, standards and \ac{Con-Ops}. Section 4 investigates \ac{5G} capabilities and studies from the \ac{UAM} application and requirements point of view. Precise navigation systems and accurate \ac{DAA} systems, which are considered one of the key technology enablers of \ac{UAM}, are studied in Section 5. In Section 6, we investigated the satellite support for \ac{BLOS} scenarios (or multi-link support scenarios). Specifically, \ac{LEO} satellites are presented as they have a potential of high data rate and low latency support for terrestrial platforms. Conclusions and potential study items are presented in Section 7. 

\vspace{\baselineskip}

\Large\textrm{2- UAM Connectivity Challenges and Requirements}\\ \normalsize


\Ac{UAM} transportation is aiming to provide services for airborne passengers, and cargo applications. The ideal world would be an autonomous vehicle that could reliably provide the service, including assurance of safe operation for humans on board. The system requirements need to consider the challenges and differences between urban and rural environments, i.e., dense urban settings like New York city etc. These kinds of urban environments imply greater challenges than in existing aviation, given the density of \ac{UAM} aircraft will be higher, and \ac{LOS}, \ac{NLOS}, and \ac{BLOS} links will experience intermittent blockages. In addition, any navigation signaling will also be stressed in terms of accuracy and latency. Given the likes of terrestrial services such as UBER, Lyft and others, the application that is drawing the most attention is the concept of air taxi. This in turn is driven by the recent advances in autonomous vehicles, lead by Tesla, which include battery technology, battery powered propulsion systems, and communication and navigation systems. The success of such services will highly depend on system robustness and safe operations.

Several, if not all, elements of \ac{UAM} are under investigation: propulsion systems, flight control systems, precise and rapid navigation and surveillance, emergency landing systems and terrain mapping, use of \ac{AI} for enhanced flight safety, off loading processing using cloud services, \ac{C2} and \ac{nC2} (sometimes referred as payload) communications, detect, sense, and avoid, and related sensors. Given the operational reliability of these sub-systems, the possibility of unmanned \ac{UAM} vehicles managed by a \ac{RPIC} is also being considered. For such a case, the communication systems must be extremely robust, secure, always available and should satisfy a high level of data integrity. While independently some or most of these systems \textit{do exist}, the need for low size, weight, power, and cost to meet the business targets are key drivers.

\subsection{Data Types: C2 and non-C2}

The definition of \ac{C2} data is all information that is involved with flight controls or safety systems. This is also evolving given how much of flight system diagnostics data will be needed in real-time, to assess flight operation for possible abort. Example \ac{AI} operating across cloud services could be critical to flight operation, the use of live images to assist remote operation, etc. All other data will fall into the \ac{nC2} category, i.e., downloading of post flight data, passenger information etc.  Thus the requirements need to consider these challenges to provide such connectivity for \ac{CNPC}, also known as \ac{C2} communication. In addition, the functions of \ac{CNPC} can address different types of information such as telecommand messages, non-payload telemetry data, support for navigation aids, \ac{ATC} voice relay, air traffic services data relay, target track data, airborne weather radar \ac{DL} data, non-payload video \ac{DL} data, etc. The \ac{C2} communication link should also support secure and reliable communications between the \ac{RPIC} and the aircraft to ensure safe and effective \ac{UAV} flight operation. 

Relative to \ac{C2}, the payload communication link \ac{nC2} may be used for data applications that often requires high throughput. Payload communication types depend on application (e.g., agriculture, public safety, passenger on-board), and can thus vary widely. It can be assumed that disruption of payload \ac{nC2} links —albeit inconvenient— is not critical, whereas \ac{C2} link disruption will be critical.

\subsection{Integrity}
Given the increase in cyber-malicious activities, \ac{UAM} \ac{C2} data communications will have to support several layers of data integrity and authentication. Creating a complete separate and custom communication link with protected spectrum seems like the logical path, but funding such large scale global deployment will be at a likely prohibitive cost. Thus the ability to leverage the existing commercially or crowd-funded systems is appealing. Such an integrity system will ensure messages received by the \ac{UAM} nodes are authenticated and verified. Thus the evolution of commercial \ac{5G} and \ac{LEO} satellites services are in the trade space. If and when the services evolve towards \ac{RPIC}, requirements to support such feature must be addressed comprehensively.

\subsection{Spectrum and Carrier frequencies}

The highly priced and congested spectrum is the real estate of future communication systems. Thus the cellular frequency bands (sub-$6$~GHz), \ac{3GPP} \ac{5G} \ac{NR}, \ac{mmWave} bands ($24$~GHz–$86$~GHz), evolving \ac{LEO} satellite bands, as well as existing aviation bands are in the trade space. As we move to higher frequencies, radio signals experience large free-space and tropospheric attenuation, which will limit range. While the \ac{5G} \ac{mmWave} systems will offer a higher throughput, they are mostly limited to \ac{LOS} and short range, and most likely available in urban environments. The \ac{UAM} system will have to resort to lower date rates as it travels to outer perimeters of cities. This is where the insertion of satellites based \Ac{BLOS} capability  becomes viable, putting pressure for all flight services to adjust for data rates and associated delays.

\subsection{Robust/Reliable Communications}

   One obvious difference between \ac{UAM} and traditional aviation settings is the proximity of aircraft to obstacles, due to the low altitude of near-urban and within-urban environments. This poses a propagation challenge, typically termed obstruction or blockage in the satellite community, and  shadowing in the terrestrial communications community. Attenuation due to such blockages can be several tens of dB, which can be enough to sever a link. Thus connectivity between aircraft and \ac{C2} stations will most likely be accomplished with $multiple$ links. 
   
   For \ac{LOS} links, air-ground connectivity will generally be preferred, because of its power efficiency and low latency in comparison to satellite or high altitude platform system (HAPS) links. The term \ac{BLOS} is used here to mean a link that is beyond the reach of a \ac{GS}, but can be reached via other means, such as other airborne platforms or satellites. The term \ac{NLOS} is used to mean a link where no \ac{LOS} exists between the aircraft and any station, ground, airborne, or satellite. The \ac{NLOS} case is typically the most challenging, from both the power and channel distortion perspective (due to multipath components, MPCs). Once \ac{UAM} is fully deployed, \ac{NLOS} cases should also be the most rare.

From the PHY and data link perspective, the primary differences between \ac{C2} and \ac{nC2} links are as follows: \begin{itemize}

\item \ac{C2} links require extremely high reliability and availability, typically availability of $99.999\%$ or larger. Reliability/availability requirements for \ac{nC2} links vary with application, but will not be as stringent.
\item Although not all requirements have been defined, most \ac{C2} link data rates will be moderate, e.g., $<300$~kbps for compressed video transmission. In contrast, the \ac{nC2} link data rates could be orders of magnitude larger, particularly for passenger applications. 
\item The \ac{C2} link latency will need to be small for certain functions. This will require message or packet size to be very small as well when \ac{C2} link data rates are low. 
\end{itemize}

At the network layer, notable points include the following:
\begin{itemize}
\item Some \ac{C2} transmissions will be multicast and/or broadcast (e.g., for a GPS local area augmentation system), whereas many \ac{nC2} links will be point-to-point.
\item Some \ac{C2} transmissions may not tolerate the latency of relaying over multiple hops.

\item Another consideration distinguishing the \ac{C2} and \ac{nC2} links is robustness or resilience. 
\end{itemize}

In short, \ac{C2} links should be extremely tolerant of interference, and should re-establish themselves very rapidly after any disruption.

In military systems this comes under the topic of \ac{EW}. This includes for example, 
\begin{itemize}
\item Jamming, which is the intentional transmission of in-band interference that degrades performance, possibly to the point of link outage;
\item Control channel “flooding,” which is the intentional transmission of signals over multiple access channels that prevents legitimate platforms from accessing the available channels;
\item Spoofing, which is when an unauthorized entity “masquerades” as a legitimate system user or controller, and sends messages with the intent to disrupt system operation.
\end{itemize}

These three forms of \ac{EW} are listed from the least sophisticated to the most sophisticated. Depending on the resources available to the disrupting entity, adapting mobile networks of such jammers/flooders/spoofers can be deployed, which could effectively disable a network of \ac{UAM} links entirely. 

   Methods to make \ac{C2} signals robust against such disruptions include spread spectrum and multi-band signaling, signal spatial filtering, usually via beamforming and beamtracking, strong \ac{FEC} techniques, active interference cancelling, and strong encryption. Adaptive routing at the network layer can also enhance robustness, at the expense of some latency. No current commercial system uses all of these techniques; existing cellular systems employ a few, e.g., encryption and \ac{FEC}.

\subsection{Security}

Security can be discussed in terms of system robustness to disruption, which is related to availability, discussed in the previous section. It can also be discussed in terms of accuracy of message transfer (e.g., resilience to spoofing). Having {\em any} entity receive and understand the majority of \ac{UAM} \ac{C2} messages will not generally pose a problem, unless that entity uses those messages against \ac{UAM} communications (e.g., via flooding or spoofing). Anti-spoofing requires strong authentication procedures.

Security issues are highly-related to regulations for \acp{UAV}. The \ac{FAA} stated remote identification (remote ID) of \ac{UAS}, crucial to integration efforts \cite{faa_220}. As defined by \ac{FAA}, {\em remote ID is the ability of a \ac{UAS} in flight to provide identification information that can be received by other parties}. This adds more regulations, which could assist the FAA, law enforcement, and other organizations including federal agencies to track \acp{UAV}, and recognize as \ac{FAA} states “unsafe manner” \ac{UAS}, including those not allowed to fly. According to \cite{faa_220}, the new rules would facilitate the collection and storage of certain data such as identity, location, and altitude regarding \acp{UAV} as well as its control station. In June 2017, \ac{FAA} charted a comprehensive report, \ac{UAS} identification and tracking \ac{ARC} \cite{faa_2017} related to the remote ID, and mentioned technologies available to identify and track drones in flight and other associated issues. According to \cite{faa_2017}, among viable technologies mentioned are \ac{ADS-B}, low-power direct RF (such as Bluetooth, Wi-Fi, RFID), cellular systems, satellite, software-based flight notification with telemetry, unlicensed integrated \ac{C2}, physical indicator, and visual light encoding. In tables provided in \cite{faa_2017}, there are analysis and comparisons of these viable technology solutions in terms of different factors such as ease of compliance for owner/operator, performance against requirements, security, inter-operability, costs, and, etc. Although, the \ac{FAA} has indicated that any solution for remote ID and tracking should avoid causing interference on the FAA’s \ac{SSR}, \ac{ACAS}, and \ac{ADS-B} systems. Therefore, it is highly recommended that any proposal for using \ac{ADS-B} frequencies must be thoroughly investigated.

As mentioned in \cite{faa_2017} this new tracking system is also in the same direction as \ac{ATC}, or possibly \ac{ATM} in aviation. \ac{ATC} as described by \ac{FAA} \cite{faa_tmo}, has the primary responsibility of the separation of the aircraft. Controllers from traffic controllers on the airport speak directly with pilots, notifying them of traffic or weather in their vicinity, while \ac{ATM} as a whole managing system facilitate the approach to managing traffic that considers the impact of individual actions. There are many factors, for example bad weather, traffic overloads, or emergencies that requires consideration of who or what may be impacted by events, and a coordinated mitigation effort to ensure safety, efficiency and equity in the delivery of air traffic services. Without a coordinated management, flight delays due to small disruptions can quickly expand across the country, causing flight cancellations, and significant delays \cite{faa_tmo}. According to \cite{faa_2017} (the reader is referred to Section 6.6), there are recommendations on how \ac{ATC} could inter-operate and maximize the benefits of ID and tracking. \ac{ATC} operations at low altitudes especially near the major airports are often complex. As a result, \ac{ATC} does not have the capabilities to track and manage all \ac{UAS} operations, especially smaller \ac{UAS} operating at low altitudes. Therefore, the \ac{ARC} recommends the \ac{UAS} ID and tracking system should inter-operate with the \ac{ATC} automation, such that target information from the ID and tracking ground system, including ID and position, can be send to \ac{ATC}.

\subsection{Communication Zones}
\ac{UAM} aircraft requiring links with larger data rates and longer communication distances bring new research dimensions for future aeronautical communication systems: 
\begin{itemize}
    \item Power efficiency and spectral efficiency are not usually equally important for \ac{UAM} scenarios as compared to terrestrial communications. The aviation communication zones are typically larger than in terrestrial communication systems, in order to achieve longer distances. Hence, power is considered to be much more precious than bandwidth. 
    \item In aeronautical stations, power must be rigorously conserved since all of it is supplied on-board and must be shared among many power consuming systems.
    \item Aeronautical channels can often be modeled as Ricean fading channels, where the Rice factor (i.e., $K$-factor) can span a range between best case \ac{AWGN} to a near-worst case Rayleigh fading channel. 
\end{itemize}

\subsection{Availability}

The cellular industry is also interested in using \acp{UAV} to expand their capacity and revenue to provide cost effective wireless connectivity for devices without coverage by the existing infrastructure. Additional cellular applications, e.g., as user equipment or relays, are also likely.

\subsection{Network Topology: Distributed or Centralized}

 Multiple links can be provided in centralized coordinated multi-point transmission and reception schemes through 5G-like techniques and systems. Multiple links can also be provided by mobile ad-hoc networks; these generally require more advanced procedures at the network layer, e.g., adaptive routing. A third option for multiple links can be satellite communications or \ac{HAPS}.

\subsection{Date rate/Bandwidth}

Data rates for \ac{C2} links are expected to be modest (e.g., a maximum of $300$~kbps for compressed video, which would not be used continuously). In contrast, the payload communication (\ac{nC2}) link is usually used for data applications, and often requires high throughput. Passengers on \ac{UAM} aircraft will desire (or require) data rates similar data to those that they obtain in terrestrial networks.  

\subsection{Air-Interface Design Considerations}

Power efficiency is of critical importance in a \ac{UAM} scenario, compared to emphasis on spectral efficiency in terrestrial networks. As cell sizes increase, the amount of transmit \ac{PA} back-off becomes a very critical issue. For \ac{UAM} scenarios, power-efficient constant-envelope modulation schemes may need to be considered.

One critical technical drawback of new generation aeronautical waveforms (\ac{LDACS}, \ac{AeroMACS}, i.e., in general multicarrier (i.e., \ac{OFDM}-like) schemes) is that these waveforms have a high \ac{PAPR}. Therefore, in order not to overdrive the \ac{PA}, a back-off of around $12$-$18$~dB is needed from the maximum output power of the PA. In a one-way communication link-budget, this corresponds to a range degradation: e.g., $12$~dB back-off means a factor of $4$ smaller range; $18$~dB back-off means a factor of $8$ smaller range. This is in comparison to traditional single-carrier schemes such as the \ac{VDL} waveforms. Roughly, cell sizes around $300$~km in \ac{VDL} reduce to $37.5$~km to $75$~km. 

On the other hand, \ac{VDL}-like waveforms have a drawback in terms of limited bandwidth (data-rate) (i.e., \ac{VDL}~Mode~2 or Mode~3 bandwidth is $25$~kHz). The aeronautical community is more cautious about moving to OFDM-based waveforms as the communication ranges are larger in aeronautical scenarios and deployment costs will increase dramatically (as more cells will be needed to cover a given area). Yet, similar problems in the power requirements of terrestrial cellular have been studied and special waveforms (i.e., \ac{DFT-s-OFDM}) that decrease back-off are already being investigated in the cellular communications community. The use of multicarrier schemes was primarily driven by two reasons: their relatively simple equalization for dispersive terrestrial channels, and the medium access flexibility provided by joint time-frequency partitioning. Aeronautical channels with strong \ac{LOS} will not experience significant dispersion for short-range \ac{UAM} links, but the flexibility of multicarrier schemes is still desirable. In addition, even if range for some \ac{UAM} links is short, the small platform size of some aircraft will still translate to power-limited communication links.

As \ac{VDL} signals have near constant envelope by virtue of their single carrier design, and hardware presently exists to accommodate preliminary phases of \ac{UAM}, the aeronautical community may employ \ac{VDL} type waveforms for the initial \ac{C2} applications. As the market grows and additional  requirements become important, the \ac{UAM} community should consider new waveforms, perhaps with the support of the \ac{5G} community. These waveforms should provide some multipath resilience for a Ricean aeronautical channel and a low \ac{PAPR}, while maintaining availability, security, integrity, and data rate requirements. The developments related to \ac{3GPP} \ac{5G} \ac{NR} for \ac{NTN} are summarized in Section 4.

\vspace{\baselineskip}

\Large\textrm{3- UAM Programs: NASA and RTCA Activities}\\ \normalsize 


The \ac{UAM} \ac{Con-Ops} will notionally consist of the following:
\begin{itemize}
    \item Shared airspace: \ac{UAM} traffic will coexist with other aircraft.
    \item \ac{UTM}: When more than one vehicle operates within a defined airspace, a traffic management system must be established to manage the airspace and disseminate real-time situational awareness information to RPICs and/or pilots on board, or between the on-board flight management systems of aircraft.
    \item Air corridors: In city centers \ac{UAM} nodes will be required to travel in a designated air highway or tunnels. Here, an effective “box” or “lane” of some height and width will be designed with speed limits and rules for the air highway.
    \item Designated takeoff/landing sites: \ac{UAM} operations will only use fixed departure and arrival air terminals, with equipment service capability and passenger loading and unloading.
    \item \Ac{VTOL} only: It is likely that only vertical takeoffs and landing will be allowed in the city-center, via stations often referred to as vertiports.
\end{itemize}

\textit{A. NASA Activities} 

\begin{figure}
\centerline{\includegraphics[width =3.5in]{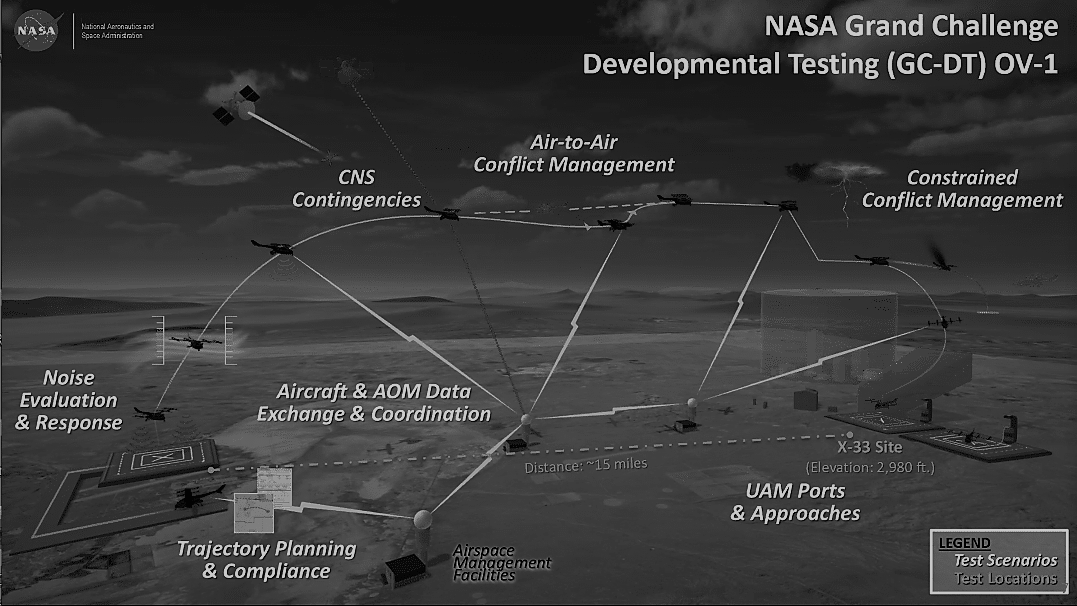}}
\caption{NASA Grand Challenge: Definitions and Scenarios\cite{NASA_2019}}
\label{fig1}
\end{figure}

NASA launched an \ac{UAM} Grand Challenge Program and the concept is illustrated in Fig. \ref{fig1}. This is a four-year testing program on unmanned aircraft integration, and beginning in $2020$, NASA will start field trials in urban environments with select participants, evaluating all subsystems of \ac{UAM} operations under a variety of weather, traffic and contingency conditions. The objective of this program is to establish an understanding of different systems and the performance required to achieve a mature \ac{UAM} system capable of operating efficiently in dense urban environments. During these test trials, NASA and government partners will conduct tests and collect data on various phases of \ac{UAM} operation to help determine vehicle certification requirements. Testing will be focused on \ac{UAM} with passenger carrying capacities, with vertical takeoff and landing and potentially standard takeoff and landing.

\textit{B. RTCA Activities} 

In USA a standard has been adopted for CNPC, created by the \ac{RTCA} \cite{RTCA_2016}. This standard specifically pertains to the L-band ($900$-$1000$~MHz) and a portion of C-band allocated to aviation ($5.03$-$5.091$~GHz). The standard applies to \ac{AG} links \ac{LOS} only, and the committee is at work on the \ac{BLOS} standard. Estimated \ac{UAV} \ac{CNPC} bandwidth requirements for the year $2030$ are $34$~MHz for the terrestrial-based \ac{LOS}  CNPC, and $56$~MHz for the satellite based \ac{BLOS} \ac{CNPC} link \cite{ITUMRBW}. The \ac{RTCA} standard does not specify any \ac{5G} applications, and primarily addresses the three lowest layers of the communications protocol stack. The standard is though general enough to be used for any type of \ac{5G} application involving medium and large sized \acp{UAV}.
In the United States, \ac{UAS} \ac{CNPC} deployment is planned in two phases, in which Phase~1 supports terrestrial networks (based on proprietary handover functionality) but does not address any industry standard handover capability, which will be addressed in phase 2. The frequency bands allocated for \ac{CNPC} Phase~1 are L-band and C-band. For future \ac{CNPC} \ac{BLOS} CNPC, defined in Phase 2, satellite communications using L, C, Ku, or Ku bands, as well as networked terrestrial and C-band terrestrial will be considered.

\vspace{\baselineskip}

\Large\textrm{4- 5G New Radio towards UAM}\\ \normalsize 

In \ac{3GPP} \ac{5G} \ac{NR} terminology \cite{nr_811}, \ac{NTN} are defined as networks, or segments of networks, using an airborne or space-borne vehicle to employ transmission to or from a relay node or base station. The \ac{NTN}s are expected to enhance the \ac{5G} \ac{NR} service reliability by providing service continuity for \ac{M2M} and \ac{IoT} devices or for passengers on board moving platforms such as aircraft. The standards are also aiming at ensuring service availability anywhere, especially for critical communications and future aeronautical communications. In 3GPP, to support \ac{NTN}, the potential areas that impact the specifications, e.g., the physical channels and modulation in Release 16 \cite{nr_phy_r16},  are identified as follows: 1-Propagation channel, 2- Frequency plan and channel bandwidth, 3- Power-limited link budget, 4- Cell pattern generation, 5- Propagation delay characteristics, 6- Mobility of the infrastructure's transmission equipment, 7- Service continuity between land-based \ac{5G} access and non-terrestrial based access networks, and 8- Terminal mobility.  However, it is still not clear if the near-term developments related to \ac{NTN} can address the requirements related to \ac{UAM} or not.

There are two major technical reports \cite{nr_811}, \cite{nr_821}, which have been updated based on  agreements in \ac{3GPP} meetings, for \ac{5G} \ac{NR}  to support \ac{NTN}. While \cite{nr_811} aims at collecting findings in the \ac{3GPP} RAN meetings, \cite{nr_821} consists of a set of necessary features/adaptations enabling the operation of the \ac{NR} protocol in \ac{NTN} for \ac{3GPP} Release 17 based on the observations in \cite{nr_811}. Several topics captured in \cite{nr_811} are as follows: 1) the \ac{NTN} deployment scenarios and related system parameters (e.g., architecture, altitude, orbit, etc.); 2) adaptation of the \ac{3GPP} channel models for \ac{NTN} (e.g., propagation conditions, mobility, etc.); and 3) the  areas on the \ac{NR} interface that may need further evaluations. It is worth noting that \ac{NR} prioritizes satellite access and considers \ac{UAS} and including \ac{HAPS} as a special case of \ac{NTN} with lower delay/Doppler values and variation rates \cite{nr_821}. It is expected that \ac{3GPP} \ac{5G} \ac{NR} Release 17 supports \ac{NTN}.

The \ac{5G} \ac{NR} with the support of \ac{NTN} aims at establishing a radio link between the \ac{UE} and the space/airborne platform, called service link, operating in frequency bands above and below 6 GHz. The recommended deployment scenarios are based on a handheld/IoT device (23 dBm \ac{TX} output power) or a very small aperture terminal, e.g., bus, train, vessel, aircraft (33 dBm \ac{TX} output power + 43.2 dBi \ac{TX} antenna/ 39.7 dBi \ac{RX} antenna gain). For the air/space borne vehicles, geostationary satellites, non-geostationary satellites with low earth orbiting satellites (from $600$~km up to $1500$~km), medium earth orbiting satellites (from $7000$~km up to $20000$~km), and airborne platforms (from $8$~km to $50$~km) are considered. The details on the deployment scenarios are given in Table 5.3-5.1 in \cite{nr_811}.

\ac{3GPP} recommends a comprehensive channel model which includes outdoor-to-indoor penetration loss, atmospheric absorption, rain and cloud attenuation, scintillation, Faraday rotation, time-varying Doppler shift, etc. for \ac{NTN} performance evaluations in \ac{5G} \ac{NR}. It has been noted that the fading characteristics can be modeled as mostly Ricean with a strong direct signal component, but slow fading can also occur due the temporary shadowing due to the obstacles such as trees, buildings, or other structures, e.g., large bridges. On the other hand, \ac{HAPS} may yield more multipath components than satellites. The \ac{NTN} channel characteristics can have an impact on reference signal design, particularly for random access channels, as \ac{NTN} would introduce large Doppler shifts compared to those  for terrestrial networks. Larger subcarrier spacing for OFDM (i.e., smaller symbol duration) is recommended to protect against distortion due to Doppler shifts. In \ac{5G} \ac{NR} R15, \ac{PTRS} was introduced, which was not available in 4G LTE \cite{lte_2018}. It has been noted that flexible configuration of \ac{PTRS} would be helpful to compensate residual carrier frequency offset and Doppler shifts. However, from the large scale perspective, supporting very high speed \acp{UE} such as aircraft systems featuring maximum speeds of $1000$~km/h is still challenging, and may require some faster power control loops than what \ac{5G} \ac{NR} offers.

The \ac{3GPP} recognizes the benefit of satellites for critical communications including public safety communications due to their dependability and large coverage area, while accepting the issue of the propagation delay of satellite systems for certain applications requiring low latency. It has been noted the one-way delay between the \ac{UE} and the RAN (on-board the satellite/\ac{HAPS} or on the ground) can reach up to 272.4 ms for \ac{GSO} systems and is larger than 14.2 ms for \ac{NGSO} systems. For \ac{HAPS}, the one-way delay is less than 1.6 ms, which is comparable to cellular networks. It has been emphasized that it is likely that these delays impact all signaling loops, especially at access and transport levels. 

The large coverage also creates some issues, e.g., more variations on the propagation delay. For example, one issue is that the ratio between propagation delays at cell center and cell edge is likely to be higher in the context of \ac{HAPS} than geostationary satellite systems, which may alter the design of contention-based access channels (of which there may be very few in \ac{UAM} links). Another issue related to the large varying delay due to the fast moving satellite/\ac{HAPS} and UEs is that the timing advances of the UEs may need to be updated more dynamically. In \cite{nr_821}, autonomous acquisition of the TA at \ac{UE} with known location and satellite ephemeris or indication of common TA to all users within the coverage of the same beam with broadcasting was considered as potential solutions. The \ac{HARQ} scheme in \ac{NR} and adaptive modulation and coding are also expected to be impacted because of large timing jitter values in \ac{NTN}. Another issue related to propagation delay is the duplexing mode. In \cite{nr_811}, it is recommended not to discard TDD for \ac{HAPS} and LEOs (in case regulations (ITU/R and/or national) allow), although \ac{TDD} requires guard times, which are a function of propagation delay, to avoid simultaneous transmit and receive.

The main design challenges in \ac{NR} are listed as 1) achieving high throughput for power-limited link budget (e.g., for a given transmit power from the \ac{UE} on the \ac{UL} and from the satellite/\ac{HAPS} on the \ac{DL}), and 2) availability of the service under deep fading situations (e.g., 20 and 30 dB in Ka band, similarly for shadowing in lower frequency bands). One way of improving throughput/power ratio is to set the operating point in the PA as close as possible to saturation point. However, this may require changes in both \ac{UL} and \ac{DL} channels in \ac{5G} \ac{NR}. For a satellite/\ac{HAPS} on the \ac{DL}, OFDM that is considered for terrestrial network may not be the best solution. For example, in \cite{nr_811}, it is emphasized that an additional 2 dB back-off can reduce the link capacity by 20\%-40\%. Finally on this topic, we note again that \ac{UAM} \ac{C2} link data rates will not be very large, so achieving high throughput is of lesser importance than {link reliability} and {availability}.

In \cite{inmarsat_2019a} and \cite{inmarsat_2019b}, Inmarsat, Intelsat, SES and Fraunhofer IIS proposed that alternative OFDM-based schemes need to be considered in the \ac{DL}, which are closer in characteristics to current state of the art \ac{SC} schemes, such as \ac{SC-FDMA} (i.e., \ac{DFT-s-OFDM}) and variants. Although this essentially requires redesign of the shared channels, i.e., a major standardization impact, it may be a necessary improvement for reliable \ac{NTN} given the large \ac{OBO} for \ac{OFDM} waveform. In \cite{nr_811}, it is also emphasized that extended multicarrier modulation and coding schemes for \ac{UL} that features low PAPR are needed. To ensure operation at low $E_s/N_0$, alternative modulation and coding schemes and redefined physical resource blocks (e.g., single tone transmission rather than 12 subcarriers) are recommended. Note that the PAPR is a function of the number of subcarriers and of the modulation order, depending on the waveform type.

\vspace{\baselineskip}

\Large\textrm{5- Navigation and Detect \& Avoid}\\ \normalsize 


Growth in the number of \ac{UAM} nodes will demand a high-accuracy positioning system from both self awareness and situational awareness points of view. Addressing sufficiently sophisticated  implementable sensor fusion and tracking algorithms to provide both accurate self-positioning and situational-awareness (what is around to avoid) has a vital importance toward reliable \ac{UAM} operations.

The FAA regulations will strictly ask UAM drones to “squawk” which means to broadcast their ID and position via drone remote ID gear as well as manned aircraft ID that used in ADS-B systems and transponders. This is essential for sense and avoid systems and is the goal of FAA and industry. By 2020, all aerial vehicles should mandatory equip ADS-B “out” system which allows vehicles to broadcast their position, vector, altitude and velocity \cite{magazine_uam_nozh11}. All manned aircrafts operating in airspace should use Mode C transponder, however, as reported in \cite{magazine_uam_nozh11} regulations excluded manned aircraft without electrical systems which as reported in \cite{magazine_uam_nozh11}, 30 percent of crop dusters in Mississippi Delta had no electrical system. Compliance issue reported in a study by FAA, that for one example they reported just six months out from mandatory ADS-B compliance only 44 percent of general aviation had installed ADS-B out equipment. Note that FAA only mandates ADS-B out equipment and not “in”.  which gives aircrafts the ability to follow position, ID of other traffic targets. FAA still not decided on equipping drones with ADS-B out because of concerns of saturating the ADS-B broadcast frequency. Note that ADS-B “in” does not have a similar problem and is under consideration for small drone mounting which makes it an open topic for small drones industry to see if customers are willing to pay for drone remote ID receivers. Author in \cite{magazine_uam_nozh11}, claimed that the drone community and FAA regulators understand this problem and are designing \ac{UTM} with drone remote ID and ADS-B “In” as an internal part of collision avoidance and sense and avoid plans.

\textit{A. Global Navigation Satellite System (GNSS) and Inertial Navigation System (INS)}

As discussed in \cite{previous_paper}, one solution is to fuse all existing \acp{GNSS} (e.g. \ac{GPS}, \ac{GALILEO}, \ac{QZSS}, \ac{GLONASS} etc.) with improved inertial navigation systems to achieve high accuracy positioning solutions. \ac{GNSS}  systems have been designed for \ac{LOS}  environments and therefore are not so suitable for moderately dense urban areas. Based on [12], the integration of all mentioned systems will result in high accuracy, multi-band and multi-mode grade receivers whose cost are not suitable for low-cost commercial applications.

Cellular networks can also play a significant navigation role in \ac{UAM} operation; specifically \ac{5G} implementations can provide node-to-node communication capabilities that lead to real time channel estimation and thus alternate navigation techniques. Based on \cite{nozh1} and \cite{nozh2}, the distribution of measured data showed standard deviation of a single mode GPS receiver to be about $2$~m, differential GPS (DGPS) $1$~m, while cellular assisted \ac{RTK} is about $16$~cm. Novel algorithm optimization and dual antennas solution such as in \cite{RTK_OPTIMIZATION}, improved the three-dimension precision of the relative positioning of \ac{RTK} under open sky condition within $2$~cm.  The accuracy situation is worse for city canyons, viaducts, and tunnels, with positioning drift of ($20$~m+) due to signal blockage and multi-path components, Also in height, when using \ac{GNSS}  alone. 
\begin{figure}
\centerline{\includegraphics[width =3.8in]{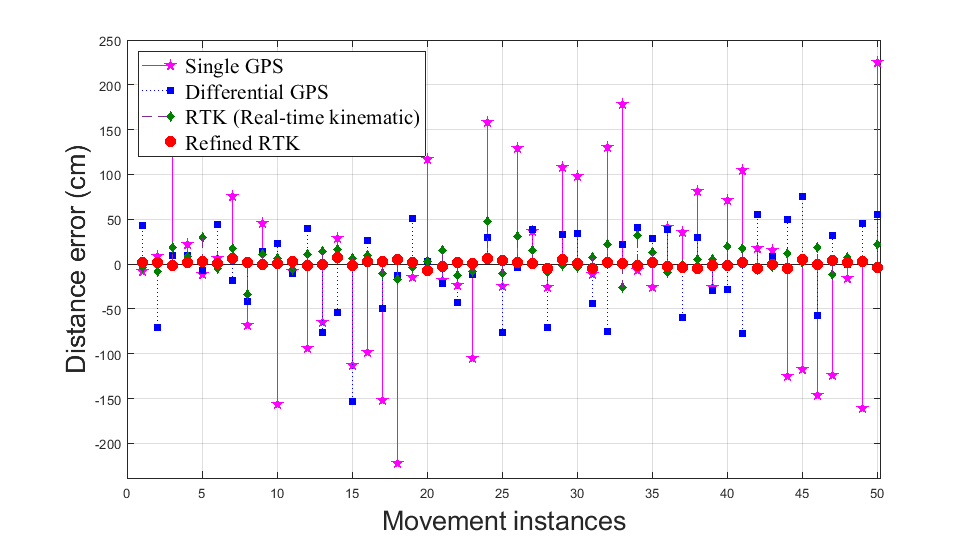}}
\caption{Comparing different positioning error distribution for different \acp{GNSS}.}
\label{figgnsscomparison}
\end{figure} Fig. \ref{figgnsscomparison} shows the positioning error distribution comparison for these four systems. This positioning variance is related to an air traffic control separation, which is the name for the concept of keeping an aircraft outside a minimum distance from another aircraft to reduce the risk of collision, as well as prevent accidents due to secondary factors, such as wake turbulence.

In \cite{peral2018methodology_nozh3} and \cite{7583720_nozh4}, exploiting \ac{5G} \ac{NR} for $450$~MHz to $6000$~MHz and $24.25$~GHz to $52.6$~GHz transmission for high-accuracy positioning is investigated as a complementary hybrid solution to \ac{GNSS} in harsh environments where an \ac{LOS} path does not exist. The authors of \cite{7583720_nozh4} evaluated positioning accuracy of six carrierless mmWave waveforms candidates (i.e., Gaussian pulse, \ac{IFFT} pulse, Gaussian-\ac{RCP}, rectangular-\ac{RCP}, sinc-\ac{RCP}, and Hann-\ac{RCP}) through \ac{IR}  and compared them with multicarrier waveforms
(i.e., \ac{OFDM}, \ac{FBMC}, and \ac{UFMC}) in terms of ranging accuracy, where they seek for very accurate positioning even in rich-scattered \ac{NLOS} fading channels. They used an energy detector with dynamic threshold which has low-computational complexity for this evaluation and provided a good ranging accuracy and not an excellent accuracy as pertain in correlation detector. They discussed these waveform candidates for vehicle positioning and \ac{D2D} communications in \ac{5G} \ac{NR} cellular networks. 
As discussed in \cite{previous_paper}, the ultimate navigation solution should be an integration of different methods, since none of the proposed systems in the literature and implementation guarantee 99.99\% availability in any location under any circumstances. 

\textit{B. Detect-and-Avoid Systems}

One criterion for integration of \ac{UAS} into urban and near-urban airspace is to use ground-based \ac{DAA} for routine \ac{UAS} terminal area operations. This can be implemented not only on the ground but also with on-board support. NASA \cite{nozh5} used Ikhana aircraft equipped with a prototype \ac{DAA} system doing $11$~flights and $200$~scripted encounters with other aircraft. The aircraft was equipped with \ac{DAA} sensor/radar, \ac{ADS-B} and traffic alert and collision avoidance system (TCAS). \ac{UAM} operations are planned to be autonomous, therefore, in a designated \ac{UAM} operation system, the \ac{DAA} process cannot be detection only. The central management system should be aware of \ac{UAS} current and future flight paths, while cooperative sharing of data through on-board sensing hardware that tracks other \ac{UAS} within range is complementary is also required. UAM central management unit will find the best route for each \ac{UAS}, which result in solving an optimization problem. 

An autonomous central management unit should process data and make decisions accordingly. An adaptive predictive control technique based on fuzzy logic was introduced in \cite{nozh5} for the supervisory layer. The authors of \cite{asep1996automatic_nozh6} proposed a fast dynamic mixed integer linear programming (MILP) optimization problem for efficient path planning of \acp{UAV} in various flight formations. The optimization problem in this paper relies on \ac{ADS-B} information and estimation of other aircraft velocity/acceleration to minimize time and energy consumption for collision-free formation flight. The solution to the problem is based on a branch and bound technique for two fixed and flexible aircraft formations.

\begin{figure}
\centerline{\includegraphics[width=3.5in]{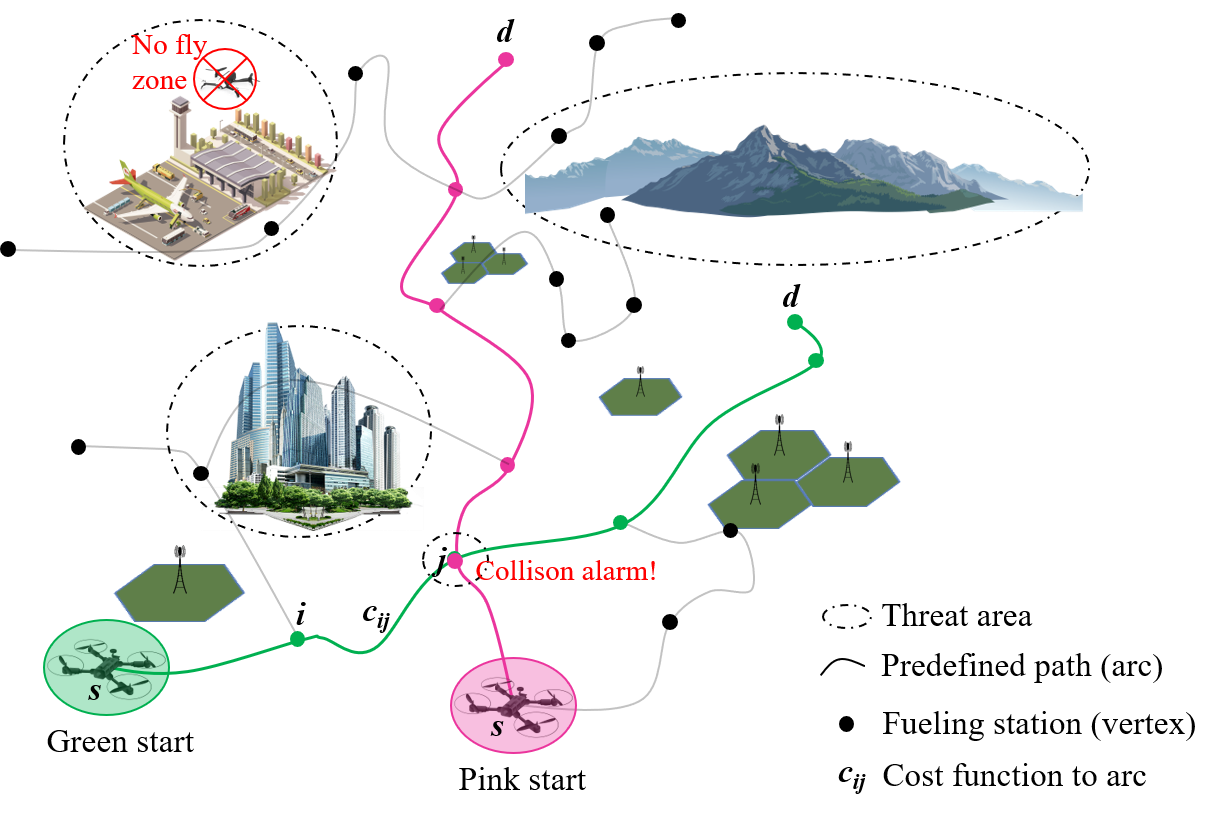}}
\caption{UAM flight trajectory problem: The minimization of the cost (which is a function of reliability of navigation, the risk associated with collision, the range of threat areas, and the location of fueling stations) for each \ac{UAV} is related to the shortest path problem and \ac{TSP} depending on the application.}
\label{figOpt}
\end{figure}

An instance of \ac{UAM} traffic management scenario is illustrated in Fig. \ref{figOpt}, where two \acp{UAV} (or vehicles for \ac{UAM}) want to reach their destinations (i.e., pink and green destinations) with minimum cost by moving along the paths between fueling stations is illustrated. We can represent this scenario as a graph where the vertices (or nodes) of the graph correspond to fueling stations. The \ac{UAV} routing scheme can be modeled as shortest path problem in graph theory. The aim is to find the minimal total cost of a tour between a starting node and destination node such that each node is visited only once. Let  $G=(V,A)$ be a directed graph, where $A$ is the set of all arcs and $V$ is the set of all nodes. Let $\delta^+ (i)$ and $\delta^- (i)$ denote a set of outgoing and incoming arcs of node $i$, respectively. In addition, let ${\rm arc}(i,j)$ be the directed path from $(i)$th vertex to $(j)$th vertex for $i,j\in V$. Let $c_{ij}$ be the cost associated with ${\rm arc}(i,j)$ represented as ($X_{ij}$). In general, $c_{ij}$ is a function of  reliability of navigation, the risk associated with collision (e.g., two \acp{UAV} reach to the same fueling station at the same time), and the range of threat areas (e.g., buildings, mountains, airports, or other no-fly zones, etc.). Let $T_{s,d}$ be a sequence defined as $({\rm arc}(s,n_1),{\rm arc}(n_1,n_2),\dots, (n_{M},d))$ for $n_{k=1,2,\dots,M},s,d\in V$, which represents a path (i.e., a tour) on the graph between the  vertex $s$ and the vertex $d$. Assuming that  $c_{ij}$ does not change in time, from a single \ac{UAV} (located at the starting vertex $s$ and heading to the destination vertex $d$) perspective, The corresponding optimization problem to find the optimum path from node $s$ to node $t$ can be given as follows:  
\begin{align}
\tilde{T}_{s,d}=&\arg\min_{T_{s,d}} \sum_{(i,j) \in A} c_{ij} X_{ij} \label{eq1a}
\\
&\textrm{s.t.,}\nonumber
\\ \nonumber
\sum_{(i,j) \in \delta^+ (i)}&X_{ij}-\sum_{(i,j) \in \delta^- (i)}X_{ji}=
\begin{cases}
1, & \textrm{if}~ i=s~ \\
-1, & \textrm{if}~ i=d \\
0, & \textrm{otherwise}  
\end{cases}
\forall i \in V
\\
\sum_{(i,j) \in \delta^+ (i)}&X_{ij} \leq 1  ~\forall i \in V
\nonumber
\end{align}
The first constraint is a flow conservation constraint which shows that the sum of the flow through arcs directed toward a node is equal to the sum of the flow through arcs directed away from that node. The second constraint  ensures that the outgoing degree of each node be one at most.

In some applications  such as mail delivery application where aircraft visit each station exactly once,  the problem transform to Hamiltonian path, where the corresponding optimization problem can be modeled as a \ac{ATSP}. The \ac{ATSP} and is \ac{NP-hard}. The reason behind this model similarity and our scenario is the very fuel dependent characteristics of \acp{UAV}. Therefore, each depot can be a fuel/re-charge station while a designated mission can be accomplished. The \ac{UAV} starts at a depot and visits a set of targets. A subset of targets can also be considered and therefore, no solution can be presented for the \ac{ATSP} in the literature, and only heuristic algorithms have been presented thus far \cite{kaplan2005approximation_nozh8}, \cite{10.1145/1383369.1383378_nozh9}, \cite{10.1145/1978782.1978791_nozh10}. The presented problem here is shown in 2D trajectories, whereas a novel ground-based aircraft separation systems are designed to follow 3D precision trajectory clearances based on navigation performance requirements (similar to Fig. \ref{figOpt}) for manned aircraft. Similar optimization modeling can be exploited for future \ac{UAM} operations. The corresponding linear integer programming formulation for \ac{ATSP} can be given by
\begin{align}
\tilde{T}=&\arg\min_{T} \sum_{(i,j) \in A} c_{ij} X_{ij} 
\\
\textrm{s.t.},& \sum_{(i,j) \in \delta^+ (i)}X_{ij}=\sum_{(i,j) \in \delta^- (i)}X_{ij}=1~, 
\label{eq2}
\end{align}
where
\begin{align}
X_{ij}=\begin{cases}
1, & \mathrm{arc}(i,j)\in \tilde{T} \\ 
0, & \mathrm{otherwise} 
\end{cases}~.\nonumber
\end{align}
The constraint \eqref{eq2} imposes that in-degree and out-degree of each vertex are equal to one. Solving this optimization problem attains the optimum sequence $\tilde{T}_{s,d}$ which minimizes the total cost. There are several methods to solve the \ac{ATSP} problem in \eqref{eq1a} with constraints in \eqref{eq2}. One solution is provided in \cite{5663700_nozh11} by adopting Lin-Kernighan-Helsgaun heuristic (i.e., one of the common approaches for solving the symmetric \ac{TSP} problem by swapping pairs of sub-tours to make a new tour) for transformed \ac{ATSP}. They first transformed the heterogeneous, multiple depot, multiple \ac{UAV} routing problem--which is a set of heterogeneous \acp{UAV} that fly from depots with a set of targets in a sequence, pass once and return to their initial depot after visiting the targets--into a single \ac{ATSP} using a Noon-Bean transformation which transform an instance of generalized  \ac{TSP} to an equivalent instance of \ac{ATSP}. They showed that solutions whose costs are within 16\% of the optimum can be obtained relatively fast for routing problem of $10$~\acp{UAV} and $40$~targets. 
The authors in \cite{7311428_nozh12} presented \ac{UAS} modeling and simulation results for Houston metroplex (city with two or more major airports and a complex airspace) conducted under \ac{FAA} system safety management transformation that includes \ac{DAA}  versus see and avoid in different mission profiles, with stochastic parameters including positional variance, fuzzy conflicts, and deviation from planned or intended profiles. They modeled Houston Metroplex by using $24$~hours of radar-track data from National Offload Program  recorded from the Offload Extractor of the Sector Design and Analysis Tool. The under test \ac{DAA} system modeled as conflict detection every 2 minutes with a look ahead range of $8$~minutes. The look ahead covered all the flights in the sector using their planned/current 4D profile. When conflict was predicted with any other manned flight, the resolution logic was only applied to the \ac{UAS} and not to the manned flight. The conflict avoidance procedure starts with a speed reduction, a vector to avoid the conflict zone and then return to the original profile, and finally a lateral offset to the original planned profile until top of descent.

\textit{C. Detect-and-Avoid Using Chirp Signals}

Sense and avoid for small-\ac{UAS} with \ac{SWAP} requirements was discussed in \cite{6712628_detect_nozh33}. Their sense and avoid radar system was designed to be a multi channel, \ac{FMCW} or chirp radar. Chirp systems simultaneously transmit swept frequency signals and receive target echos. Similar approaches were used for channel characterization using chirp signals in \cite{wideband_chirp_nozh44} except that the transmitter and receiver can be located at different coordinates. As described in \cite{6712628_detect_nozh33} and \cite{wideband_chirp_nozh44}, windowing and fast-Fourier transforms are performed to obtain range (by estimating the gains of the multi-path components) and speed (by estimating Doppler shift) of the target. One challenge in \ac{DAA}  chirp systems is dealing with the leakage of signal resulting from the \ac{LOS} radiating path. 
Authors in \cite{6712628_detect_nozh33} used Analog Devices integrated circuits (i.e., ADF4158 and AD8283) for radar applications, which provide 6 channel analog-to-digital converters in their modified evaluation board. This solution was motivated by cost and safety requirements for small \acp{UAV} for the future. 

New autonomous vehicles are using \ac{ADAS} with forward-collision warning tool. These systems enable semi-self driving cars by using sensors such as \ac{LIDAR}, camera and radar. \ac{FMCW}  radars being used in ground vehicle's \ac{DAA}  systems and can be a good candidate for future \ac{UAV}  implementation. The authors in \cite{chirp_radar_sensor_nozh444} investigated \ac{FMCW}  radar-to-radar interference, which impose a ghost-target problem in cars. They proposed different chirp slope in \ac{TF} domain to reduce interference. Implementing these sensors for scenarios with high \ac{UAV}-to-\ac{UAV} collision probability enabled by novel interference mitigation techniques is suitable. In \cite{sahin_2020}, chirp signals are synthesized by using Bessel functions and Fresnel integrals as frequency-domain spectral shaping for \ac{DFT-s-OFDM} with the motivation of standard-compliance and dual radar-and-communication function, which may also be important features for \ac{UAM}.

\vspace{\baselineskip}

\Large\textrm{6- LEO Satellite Constellations}\\ \normalsize 

Advancements in reusable rocket technologies, solar cells, ion propulsion systems, and spacecraft collision avoidance systems together with precise navigation systems, phased array antenna systems and high bandwidth communication strategies open up a new era in \ac{SWAP} \ac{LEO} constellation ($250$~kg weight and $600$~km altitude) communication architectures. As the two way communication (from \ac{GS} to \ac{LEO} satellite (600 km) and back to \ac{GS}) propagation delay is about $4$~ms, this architecture enables delay sensitive, high data rate applications that were impractical with a \ac{MEO} ($2000$~km altitude) / \ac{GEO} ($36000$~km altitude) satellites (where two way propagation delays varies between $100$~ms to $250$~ms)\footnote{These values only consider the propagation delay, there will be additional processing and application delays.}.  

There are many projects targeting this service utilizing \ac{LEO} satellites: Starlink of SpaceX \cite{Starlink}, oneWeb, Facebook (PointView’s Athena) \cite{Athena}, and Kuiper of Amazon \cite{Amazon}. Among those, Starlink is targeting more number of satellites in the Northern U.S. and Canada in 2020, rapidly expanding to near global coverage of the populated world by 2021. Other projects have a plan for starting their services before 2025. When these schedules are projected on a \ac{UAM} timeline, there is a possibility that \ac{LEO} constellations could support \ac{CNS} technologies towards \ac{UAM} applications. Robustness of the technologies should be evaluated from \ac{UAM} requirements perspective as they begin to take shape in the upcoming years. 

The upcoming \ac{LEO} satellite providers are focusing on providing data links, not just to ground users, but also to serve as the backbone for remote cellular towers, where cost of installing ground cables is prohibitive. These system may address or compliment \ac{UAM} communications when the \ac{LOS} link is not available. Even though the propagation delay for \acp{LEO} is significant compared to that of terrestrial links, a single \ac{LEO} satellite can provide a single-hop link between two points providing a footprint coverage of several hundred km.

\begin{figure}
\centerline{\includegraphics[width =3.5in]{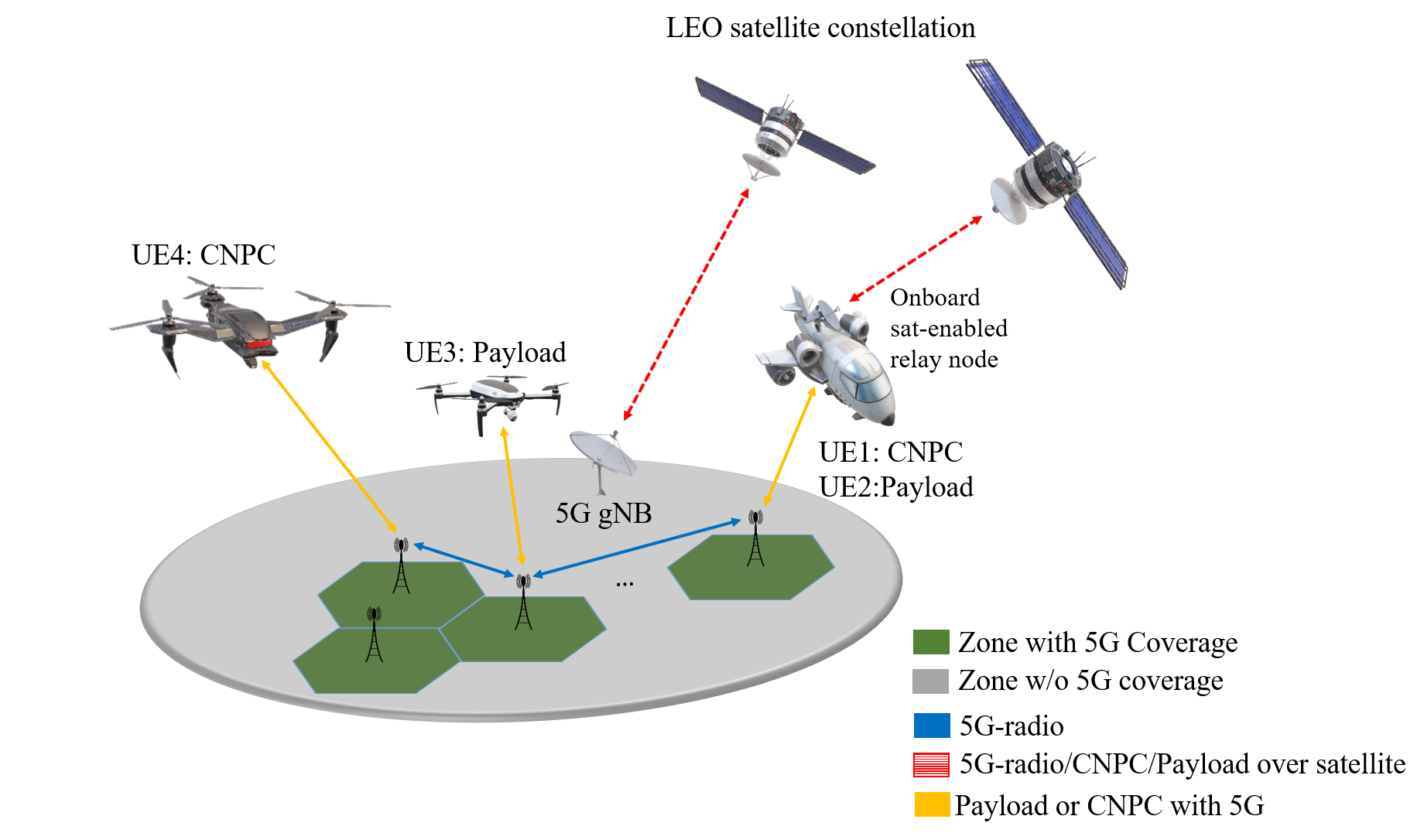}}
\caption{\ac{BLOS} \ac{CNPC} link using the \ac{5G} relay node concept: on-board \ac{5G} relay node and \ac{LEO} satellite \cite{previous_paper}}
\label{figLeo}
\end{figure}

A possible approach to smoothly integrate \ac{UAV} into the \ac{5G} systems via \ac{BLOS} links using \ac{LEO} satellites is the relay networks illustrated in Fig. \ref{figLeo}. A low-complexity satellite-enabled \ac{RN} onboard a \ac{UAV} transports the \ac{5G} \ac{DL}/\ac{UL} waveform via a \ac{LEO} link to the terrestrial base station, the so-called donor NodeB at the ground earth station. For the on-board equipment (flight controller requiring \ac{CNPC} or payload equipment requiring a high-throughput link) the RN appears like a \ac{GS}. The donor NodeB, for whom the RN is transparent, simply sees a number of users. While this approach may not be optimal in the sense of achieving channel capacity, it requires less communication infrastructure both on-board in \ac{UAM} node and at the \ac{GS}, simplifies handover from \ac{5G} to satellite, and leverages \ac{5G} technology.

\vspace{\baselineskip}

\Large\textrm{7- Conclusions}\\ \normalsize


\ac{UAM} will provide a new era for aeronautical \ac{CNS}. Critical \ac{UAM} requirements are derived from \ac{C2} (particularly in \ac{RPIC} scenarios), data connectivity for passengers and critical flight systems, \ac{UAS}-to-\ac{UAS} communication to avoid collision, and data exchange for positioning and surveillance.

Future challenges for \ac{UAM} operations are predicted to be safety and integration barriers across the entire \ac{UAM} ecosystem in highly populated zones; accurate and navigation and surveillance capabilities; interactions between vehicles for both traditional and novel \ac{UAM} airspace management systems; \ac{DAA}  capabilities; handling the loss of primary communications; and public acceptance or response to vehicles. These all represent areas of future investigation. Specifically, the following issues has to be addressed extensively.

\begin{itemize}
    \item The requirements for robust \ac{UAM} scenarios will have many updates during the development stages. To have a successful operation of \ac{UAM}, requirements to possible solutions must be addressed systematically.
    \item Quantification of the integrity, availability, security, and robustness/reliability is a key enabler of \ac{UAM}.  
    \item \ac{5G} studies and working groups are solely based on cellular companies and vendors. Requirements of aviation support must be addressed in \ac{5G} standards including \ac{UAM} applications.
    \item Dense and highly obstacle-laden \ac{UAM} environments will require diverse navigation and detect \& avoid systems compared to commercial aircraft scenarios: fusing (with extensive tracking capabilities) \ac{GNSS}, \ac{IMU}, radar(s), Satcom, collision-avoidance/surveillance broadcast information (generally studied under \ac{RTCA}), and \ac{5G} data communications to address these issues should be further investigated. 
    \item The success of \ac{UAM} will depend on the existence of practically-implementable-theoretically-provable solutions.
    High-complexity, real-time \ac{CNS} applications in a "Moores-law approaching to bound" era will lead us to implementable yet sub-optimal solutions.   
\end{itemize}

\bibliographystyle{IEEEtran}
\bibliography{uam}

\end{document}